\documentclass[twocolumn]{aastex631}

\usepackage{color}

\received{}
\revised{}
\accepted{}

\submitjournal{ApJL}

\shorttitle{AT2018lqh: A Black Hole is Born?}
\shortauthors{Tsuna, Kashiyama, Shigeyama}

\begin{document}

\title{AT2018lqh: Black Hole Born from a Rotating Star?}

\correspondingauthor{Daichi Tsuna}
\email{tsuna@resceu.s.u-tokyo.ac.jp}

\author[0000-0002-6347-3089]{Daichi Tsuna}
 \affiliation{Research Center for the Early Universe, Graduate School of Science, University of Tokyo, Bunkyo-ku, Tokyo 113-0033, Japan}
 \affiliation{Department of Physics, Graduate School of Science, University of Tokyo, Bunkyo-ku, Tokyo 113-0033, Japan}

\author[0000-0003-4299-8799]{Kazumi Kashiyama}
\affiliation{Research Center for the Early Universe, Graduate School of Science, University of Tokyo, Bunkyo-ku, Tokyo 113-0033, Japan}
\affiliation{Department of Physics, Graduate School of Science, University of Tokyo, Bunkyo-ku, Tokyo 113-0033, Japan}
\affiliation{Kavli Institute for the Physics and Mathematics of the Universe (Kavli IPMU,WPI), The University of Tokyo, Chiba 277-8582, Japan}

 \author[0000-0002-4060-5931]{Toshikazu Shigeyama}
 \affiliation{Research Center for the Early Universe, Graduate School of Science, University of Tokyo, Bunkyo-ku, Tokyo 113-0033, Japan}
 \affiliation{Department of Astronomy, School of Science, The University of Tokyo, 7-3-1 Hongo, Bunkyo-ku, Tokyo 113-0033, Japan}

\begin{abstract}
Recently an intriguing transient AT 2018lqh, with only a day-scale duration and a high luminosity of $7\times 10^{42}\ {\rm erg\ s^{-1}}$, has been discovered. While several possibilities are raised on its origin, the nature of this transient is yet to be unveiled. We propose that a black hole (BH) with $\sim 30\, M_\odot$ forming from a rotating blue supergiant can generate a transient like AT 2018lqh. We find that this scenario can consistently explain the optical/UV emission and the tentative late-time X-ray detection, as well as the radio upper limits. If super-Eddington accretion onto the nascent BH powers the X-ray emission, continued X-ray observations may be able to test the presence of an accretion disk around the BH.
\end{abstract}


\section{Introduction}

\begin{figure*}
    \centering
    \includegraphics[width=\linewidth]{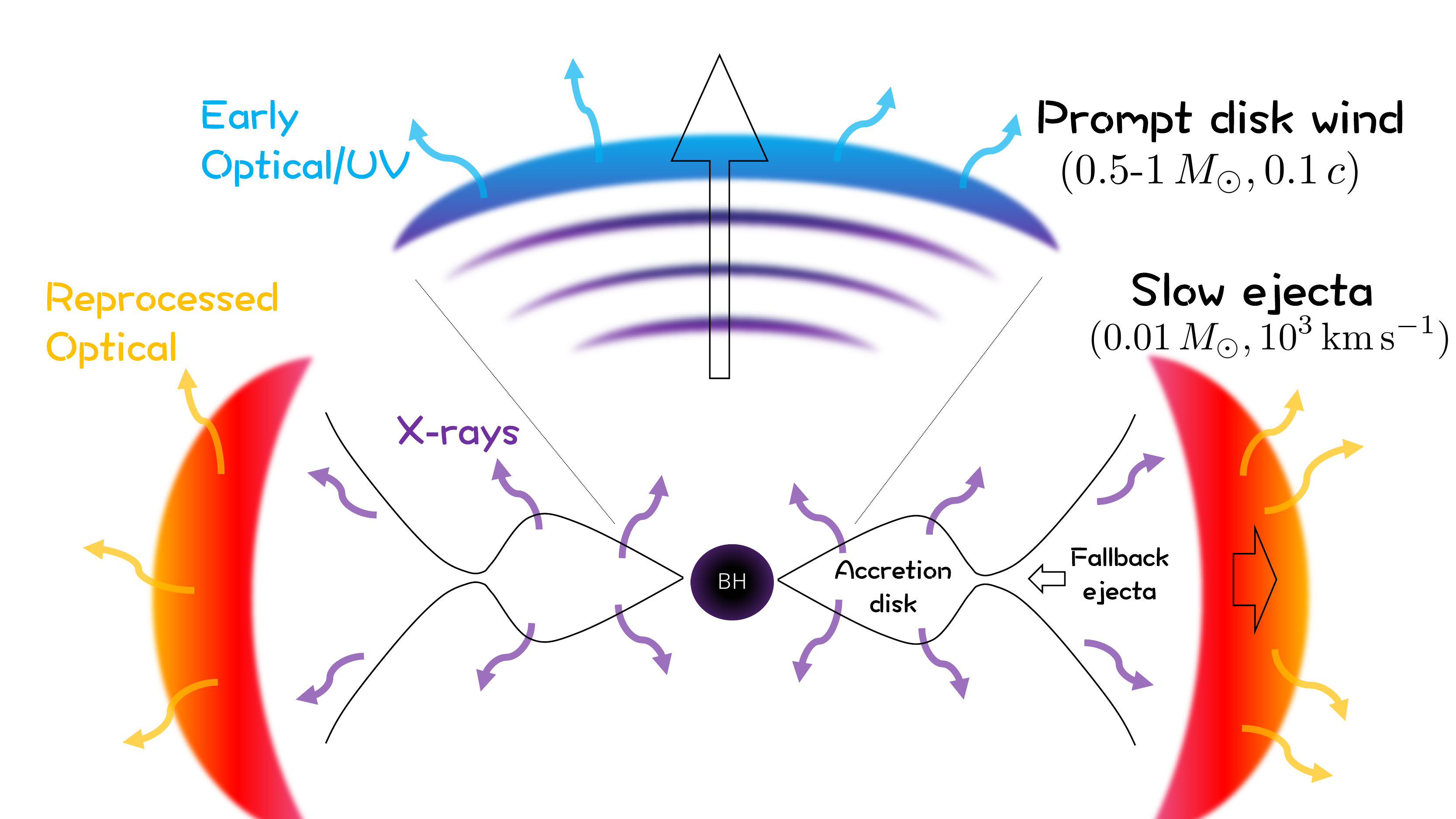}%{ponchie.png}
    \caption{Schematic picture of our model for AT2018lqh. A rotating massive star collapses to a BH, and material near the star's surface circularizes around the BH and creates a prompt accretion disk. A fast wind is launched in the polar region due to radiation pressure in the accretion disk, and becomes a source of early optical emission. Weak mass ejection due to neutrino emission in the core occurs in the equatorial region, and becomes a source of late optical emission by reprocessing the X-ray emission from the fallback disk. }
    \label{fig:ponchie}
\end{figure*}

Optical surveys are evolving in sensitivity, cadence and field of view, which resulted in finding peculiar transients that greatly differ from normal supernovae. Especially surveys with short cadence are finding a number of rapid transients with duration of order days \citep{Drout_et_al_2014,Arcavi_et_al_2016,Tanaka16,Pursiainen_et_al_2018, Ho21}, whose origin(s) is in active debate.

Recently an optical transient AT 2018lqh with a very short duration of order days was detected by ZTF \citep{Bellm19} and reported in \cite{Ofek21}. The multi-wavelength observations of AT 2018lqh can be summarized as follows\footnote{The epochs here are relative to the time $t_s={\rm JD}\ 2458310.348$, half a day before the first marginal ($2.2\sigma$) detection by ZTF.}:
\begin{itemize}
    \item The light curve evolves very rapidly, with duration above half-maximum light of 2.1 days. The peak luminosity is $\sim 7\times 10^{42}\ {\rm erg\ s^{-1}}$. A follow-up observation by Keck at day 61 detected this transient with a lower luminosity of $\approx 2\times 10^{40}\ {\rm erg\ s^{-1}}$.
    \item The color of the optical emission is found to quickly evolve from a blue color of $T_{\rm eff}\sim 2\times 10^4$ K at 1 day to $9000$ K at 3--4 days.
    \item Radio observations by VLA were carried out on day 218, and found an upper limit of $L_\nu\lesssim 1.5\times 10^{27}\ {\rm erg\ s^{-1}}$ at the C-band (5 GHz) and $L_\nu\lesssim 2.5\times 10^{27}\ {\rm erg\ s^{-1}}$ at the K-band (13 GHz).
    \item The X-ray observations by {\it Swift}-XRT on day 212 resulted in marginal detection of an X-ray source with unabsorbed luminosity $L_{\rm 0.2-10 keV}=9.8^{+9.8}_{-4.9}\times 10^{40}\ {\rm erg\ s^{-1}}$.
    \item From their archival search parameters, the event rate is estimated to be $\sim 1$\% of the supernova rate.
\end{itemize}
The optical observations require a fast ($\sim 0.1c$, where $c$ is the speed of light) material as the emitting source to power the rapid light curve, as well as some mechanism to power the late time emission at 61 days. Several possibilities are raised in \cite{Ofek21}, such as interaction with a confined CSM accompanied by shock cooling, or fast ejecta composed mostly of radioactive $^{56}$Ni that powers the light curve.

In this work, we propose that events like AT 2018lqh can occur from mass ejection when a massive star collapses to a BH. BH formation can realize mass ejection in several ways, depending on the type of the progenitor and its rotation. For core-collapse of a slowly rotating star, decrease in the core's gravity due to neutrino emission in the protoneutron star phase is expected to lead to a weak mass ejection of energy $\lesssim 10^{48}$ erg. \citep{Nadezhin80,Lovegrove13,Fernandez18}. For stars with significant rotation, infalling matter can circularize around the BH and create an accretion disk. In this case the accretion is expected to be super-Eddington, and radiation-driven wind is expected to be launched (e.g. \citealt{Ohsuga05,Sadowski14}) in addition to the weak mass ejection, which can have a large effect on the observable signatures \citep[e.g.,][]{Dexter13,Kashiyama15,Kimura2017}.

In Section \ref{sec:model} we describe our model, and compare our predictions with multi-wavelength observations by \cite{Ofek21}. We conclude in Section \ref{sec:conclusion}, raising possibilities to test our model.

\section{Our model}
\label{sec:model}

\begin{figure*}
    \centering
    \includegraphics[width=\linewidth]{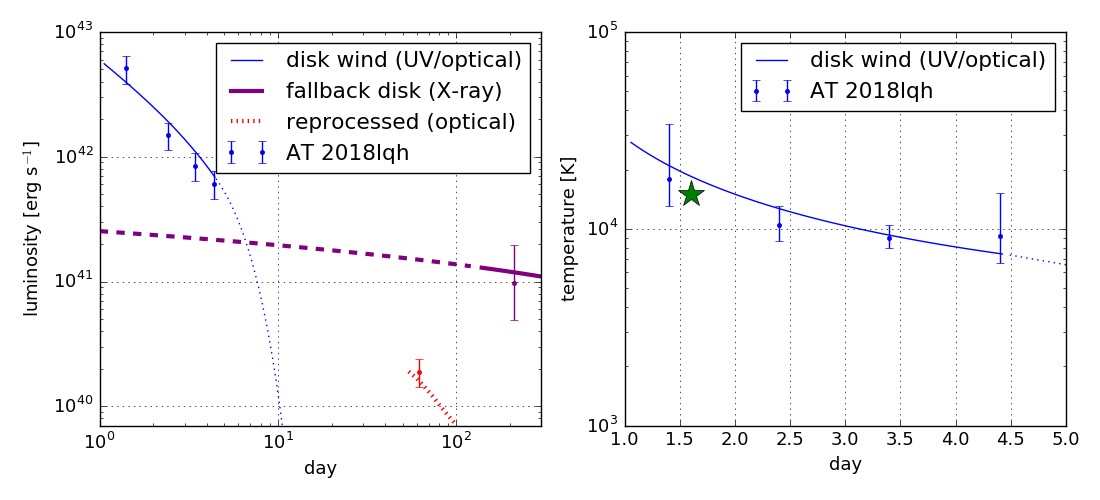}
    \caption{Luminosity and temperature evolution of the emissions predicted in our model. The points and error bars show the estimate from photometric observations of AT 2018lqh, while the green star shows the temperature obtained from spectroscopic observations \citep{Ofek21}. The thin blue line shows the cooling emission from the prompt disk wind (Sec. \ref{sec:early_optical}), whose model parameters are summarized in Table \ref{tab:optical_params}. The thick purple line shows our expectation for the X-ray luminosity from the fallback accretion disk (Sec. \ref{sec:X-ray}), calculated from the slim disk model of \cite{Watarai00} with $M_{\rm BH}=30M_\odot$. The solid and dashed rays roughly separate the time when the prompt disk wind becomes transparent to X-rays. Finally, the red dotted line shows the optical emission from reprocessing in the slow ejecta assumed to have mass $0.01M_\odot$ (Sec. \ref{sec:late_optical}). We plot only for $t>54$ days, when the optical depth of Thomson scattering inside the ejecta does not greatly exceed unity.}
    \label{fig:early_optical}
\end{figure*}

We show the schematic picture of our BH model in Figure \ref{fig:ponchie}. As discussed below, we found that a rotating blue supergiant (BSG) star with a mass of $M_* \gtrsim 30\,M_\odot$, undergoing core-collapse and forming a BH at its center, best reproduces the multi-wavelength signal.

While most of the stellar envelope sinks into the BH, the outer layer can be ejected due to neutrino mass loss at the core and/or circularize around the BH to form an accretion disk. The former corresponds to the slow ejecta, and the latter results in the prompt disk formation close to the innermost stable circular orbit (ISCO) around the BH. 

Mass ejection due to neutrino mass loss for BSGs has recently been studied \citep{Fernandez18,Tsuna20,Ivanov21}. \cite{Fernandez18} and \cite{Tsuna20} estimated that mass ejection of $0.05$--$0.1M_\odot$ can occur, while \cite{Ivanov21} investigated the dependence on the equation of state (EOS) at the core and found that a softer EOS would reduce the ejected mass down to $\sim 0.01M_\odot$. These studies agree that the ejecta expand slowly with a bulk velocity $\lesssim 10^3{\rm km \ s^{-1}}$, on the same order of the escape velocity at the BSG surface. If the progenitor is rotating, matter will be preferentially ejected around the equatorial plane due to centrifugal support.

Stellar evolution calculations suggest that a BSG collapsar can typically maintain significant rotation at the outermost layers of a few to 10 $M_\odot$ so that to form an accretion disk~\citep[e.g.,][]{Kashiyama18}, although angular momentum transport inside the star may complicate this picture (e.g. \citealt{Fuller19}). A single massive star could collapse as a rotating BSG for initial metallicity of order 10\% of solar or lower \citep{Kashiyama18}. Alternatively, binary evolution may produce such collapsars through e.g., angular momentum transfer by dynamical tides \citep[e.g.,][]{Woosley12} or stellar merger in the late phase of stellar evolution \citep[e.g.,][]{Podsiadlowski92}.

The prompt disk formation occurs at the free-fall time at the stellar surface
\begin{equation}
    t_{\rm ff}\approx \sqrt{\frac{R_*^3}{GM_{\rm BH}}} \sim 1\ {\rm day}\left(\frac{R_*}{50R_\odot}\right)^{3/2}\left(\frac{M_{\rm BH}}{30M_\odot}\right)^{-1/2}
 \end{equation}
where $R_*$ is the stellar radius, $M_{\rm BH}$ is the BH mass and $G$ is the gravitational constant. A fraction of the prompt disk mass up to a few 10 \% will be ejected by the radiation-driven wind with an escape velocity at the circulation radius \citep{Kashiyama15}. This wind is preferentially launched in the polar regions where it is not impeded by the accretion disk. A fraction $f_\Omega\sim H/r(<1)$ of the $4\pi$ steradians is subtended by the disk, where $H/r$ is the normalized scale height of the disk assumed to be independent of radius. In this model we expect that a similar fraction $f_\Omega$ around the equatorial plane is covered by the slow ejecta, while the disk wind eventually becomes close to spherical as it expands.

\subsection{Emission from the Disk Wind}
\label{sec:early_optical}

The disk wind is initially radiation-dominated when it is launched. As the wind expands, it converts most of the internal energy to kinetic energy while the rest escapes as radiation. \cite{Kashiyama15} predicts this cooling emission to be fast and luminous, as is the case for AT 2018lqh. Here we adopt their formulation to calculate the evolution of luminosity and temperature, and compare with early optical/UV observations of AT 2018lqh. At $t>t_{\rm ff}$, the wind is assumed to expand homologously with a profile
\begin{eqnarray}
\rho(r,t) &\propto & t^{-3}(r/t)^{-\xi}\\
T(r,t) &\propto& t^{-1}(r/t)^{-\xi/3}
\end{eqnarray}
where $v=r/t$ has minimum and maximum values $v_{\rm min}$ and $v_{\rm max}$, and the proportionality factors are determined as a function of $R_*$, $M_{\rm BH}$, wind mass $M_{\rm w}$ and launching radius \citep{Kashiyama15}. At each $t$ we obtain the radius $r_{\rm dif}$ where the diffusion time is comparable to $t$ by solving the equations \citep{Kisaka15}
\begin{eqnarray}
t_{\rm dif} &=& \frac{\Delta r}{c}\int^{v_{\rm max}t}_{r_{\rm dif}}\kappa\rho dr \\
\Delta r &=& v_{\rm max}t-r_{\rm dif},
\end{eqnarray}
where $\kappa$ is the opacity assumed to be a constant. We define the emission with temperature $T_{\rm obs}$ being that at $r=r_{\rm dif}$, and luminosity $L_{\rm obs}=4\pi aT_{\rm obs}^4r_{\rm dif}^2\Delta r/t$, where $a$ is the radiation constant.

In Figure \ref{fig:early_optical} we show a fit to the light curve obtained by \cite{Ofek21}. We adopted the parameters listed in Table \ref{tab:optical_params}. We find that cooling emission from a prompt disk wind with total mass of $\sim 0.6\mbox{-}0.8\,M_\odot$ moving at a velocity $\approx 0.1\,c$ can explain the early optical observations of AT 2018lqh. The inferred mass and velocity are consistent with a few $M_\odot$ prompt disk formation of a BSG collapsar. The slope of $L_{\rm obs}$ requires $\xi \approx 0.7$, i.e. the faster part of the wind carries more energy. The profile would depend on the initial angular momentum distribution in the progenitor that shapes the structure of the accretion disk, as well as dissipation and transfer of energy inside the disk. A detailed investigation of these processes is beyond the scope of this work.

Both the luminosity and temperature declines exponentially after the photon diffusion radius reaches the tail of the prompt disk wind, i.e., $r_{\rm dif}=v_{\rm min}t$. For Figure  \ref{fig:early_optical} we adopted a Gaussian decay in the light curve expected from analytical modelling \citep{Arnett80}, but we find that the best-fit parameters are not greatly affected by the exact functional form at this regime. The emission detected by Keck on day 61 requires a different interpretation, which we discuss in Section \ref{sec:late_optical}.

While similar disk winds may be launched for other progenitors like Wolf-Rayet stars (WRs) and red supergiants (RSGs), we expect that the resulting light curves would be different from AT 2018lqh. For WRs, the luminosity of the cooling emission is predicted to be an order of magnitude dimmer than BSGs, due to its smaller $R_*$ and shorter $t_{\rm ff}$ (see equation 27, 28 of \citealt{Kashiyama15}). For RSGs a significant fraction of the hydrogen envelope is expected to be ejected by neutrino mass-loss \citep{Fernandez18}, and the emission, powered instead by interaction of the wind and ejecta  \citep{Dexter13}, would be longer and possibly brighter.

\begin{table}[]
    \centering
    \begin{tabular}{ccc}
         \hline
         Parameters & Values & Range\\
         \hline
         Stellar radius ($R_*$) & $3\times 10^{12}$ cm & $(2-3)\times 10^{12}$ cm \\
         Black hole mass ($M_{\rm BH}$) & $30\ M_\odot$ & $20-30\ M_\odot$\\
         Prompt disk wind mass ($M_{\rm w}$) & $0.8\ M_\odot$ & $0.6-0.8\ M_\odot$ \\
         Launching radius ($r_l$) & $70r_{\rm Sch}$ & $(50-80)r_{\rm Sch}$ \\
         Opacity ($\kappa$) & $0.2$ cm$^{2}$ g$^{-1}$ & $0.2-0.4$ cm$^{2}$ g$^{-1}$\\
         Minimum velocity & $0.7\bar{v}_{\rm out}$ \\
         Maximum velocity &  $1.4\bar{v}_{\rm out}$ \\
         Density profile ($\xi$) & 0.7
    \end{tabular}
    \caption{Model parameters that explain the early-phase optical emission of AT 2018lqh. $r_{\rm Sch}=2GM_{\rm BH}/c^2$ is the Schwarzschild radius of the BH, and $\bar{v}_{\rm out}=c/\sqrt{70}\approx 0.12c$ is the typical velocity of the wind. See \cite{Kashiyama15} for details of each parameter. The second column shows the parameter set that reproduces the observed data points with no mismatch (see Figure \ref{fig:early_optical}). The last column shows the rough range of each key parameter that results in optical/UV predictions of at most 1 mismatch, being less than two standard deviations, when the other parameters are fixed to values in the second column.}
    \label{tab:optical_params}
\end{table}

Radio emission is expected to be produced at the collisionless shock that forms when the prompt disk wind sweeps the circumstellar material (CSM). We investigate the consistency of this model and radio observations at day 218. We use the formulations in \cite{Kashiyama18} to calculate the radio light curves, and adopt the parameters of the disk wind inferred from the fit to the optical emission. Parameters of the CSM and microphysical parameters are taken to be same as those adopted in \cite{Kashiyama18} (see Table 1 of their paper).

\begin{figure}
    \centering
    \includegraphics[width=\linewidth]{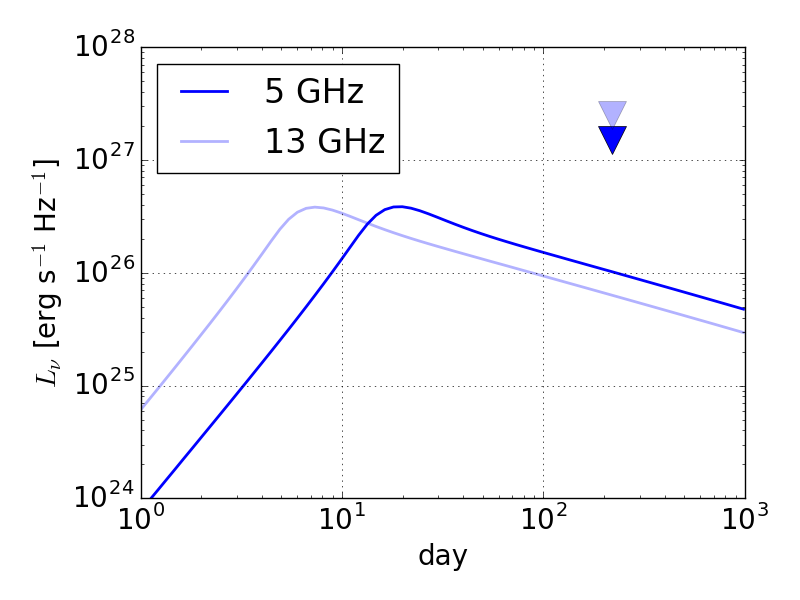}
    \caption{Radio luminosity from the prompt disk wind predicted by our model. The triangles are upper limits from the VLA observation at day 218 by \cite{Ofek21} .}
    \label{fig:wind_radio}
\end{figure}

A comparison of our radio light curve and the upper limit at day 218 is shown in Figure \ref{fig:wind_radio}. We find that the radio emission is consistent with the upper limits for both the C-band (5 GHz) and K-band (13 GHz). If these events occur nearby ($\lesssim 100$ Mpc), early radio observations may be able to detect these fast wind.

\subsection{X-rays from the Central Black Hole}
\label{sec:X-ray}
The presence of a radio point source near AT 2018lqh at a similar epoch makes it ambiguous whether the X-rays detected by {\it Swift}-XRT are associated to AT 2018lqh. If the X-rays are really associated, one may attribute them to interaction with the CSM, as observed in Type IIn supernovae (for a review see \citealt{Chandra18}). However, a dense CSM would affect the optical signatures at earlier phases. In fact \cite{Ofek21} disfavors the presence of an extended dense CSM from the absence of intermediate-width emission lines in the early phase spectra.

In our model we expect that the X-ray emission instead comes from the center, i.e. from the newborn BH and its accretion disk. Explaining the X-ray emission requires the system to be super-Eddington, with X-ray luminosity of at least 10 times the Eddington luminiosity of a stellar-mass BH $L_{\rm Edd}=2.5\times 10^{39}{\rm erg\ s^{-1}}(M_{\rm BH}/10M_\odot)(\kappa/0.2{\rm cm^{2}\ g^{-1}})^{-1}$. This may be probably around the limit that can be achieved by super-Eddington accretion onto a stellar-mass BH (e.g. \citealt{Watarai00,Ohsuga05,Poutanen07,King08}). Such luminosities are observed in a subclass of ultraluminous X-ray sources (ULXs), called hyperluminous X-ray sources \citep{Matsumoto03,Gao03}, although whether these sources are from stellar-mass BHs is uncertain (e.g. \citealt{Bachetti14,Kaaret17,Barrows19}).

As the viscous timescale in the accretion disk is much shorter than 200 days, a necessary condition is that the accretion rate onto the BH is still super-Eddington at this epoch. This can be supplied by the fallback from the slow ejecta, because the typical energy budget for the weak mass ejection ($\sim 10^{48}$ erg; \citealt{Fernandez18}) is smaller than the binding energy of the BSG envelope ($\sim 10^{49}$ erg) and a fraction of the ejecta will be bound. \cite{Fernandez18} finds through simulations and analytical arguments that for radiative progenitors (BSGs, Wolf-Rayet stars), the accretion rate at late phase follows the standard $\dot{M}\propto t^{-5/3}$ expected from the marginally bound ejecta. The proportionality factor will depend on the progenitor, but for the two BSGs chosen by \cite{Fernandez18} it is roughly $\dot{M}\sim 10^{-8}M_\odot\ {\rm s^{-1}}(t/10^6\ {\rm sec})^{-5/3}$. Extrapolating this to $200$ days and assuming that only a fraction $f_\Omega$ of the material near the equatorial plane can actually fall back, we find $\dot{M}\sim (3f_{\Omega})\times 10^{-3}M_\odot\ {\rm yr^{-1}}$, yet orders of magnitude larger than the Eddington rate. In the left panel of Figure \ref{fig:early_optical}, we plot the luminosity as a function of $t$ for $M_{\rm BH}=30M_\odot$, using the relation between luminosity and $\dot{M}$ found by \cite{Watarai00}. Assuming most of this luminosity is emitted in X-rays, we find it to be consistent with that observed on day 212. 

\begin{figure*}
\centering
\begin{tabular}{cc}
\begin{minipage}{0.5\hsize}
\centering
\includegraphics[width=\linewidth]{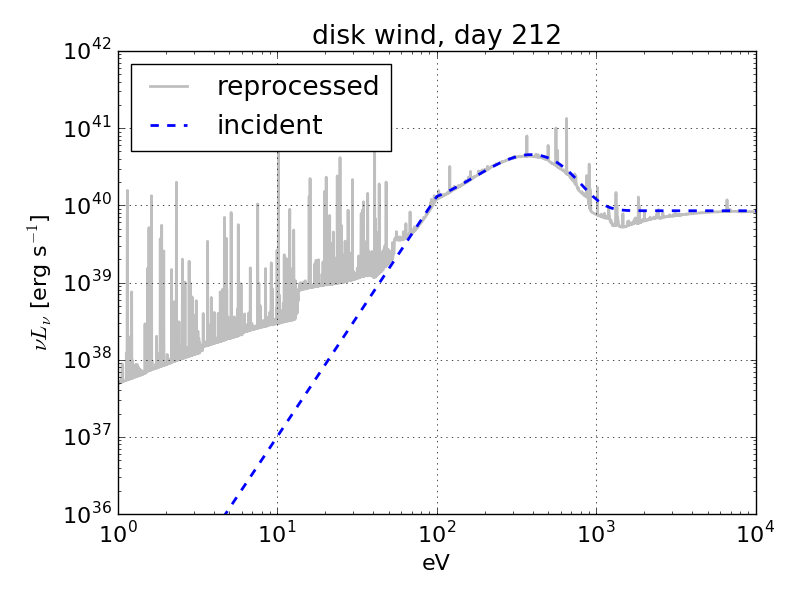}
\end{minipage}
\begin{minipage}{0.5\hsize}
\centering
\includegraphics[width=\linewidth]{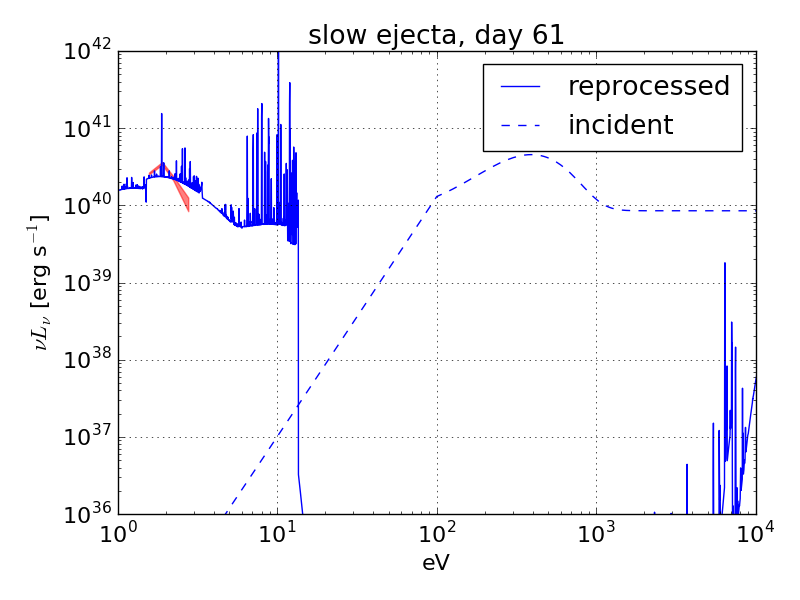}
\end{minipage}
\end{tabular}
\caption{Incident and reprocessed spectra for X-rays passing through the prompt disk wind on day 212 (left panel) and slow ejecta on day 61 (right panel). (Left panel) We assume the wind has a uniform density profile with mass $0.8M_\odot$ and outermost radius $7\times 10^{16}$ cm. (Right panel) We assume the ejecta has a uniform density profile with mass $0.01M_\odot$ and outermost radius $5\times 10^{14}$ cm. The red shaded region shows the observed flux in {\it BVRI} bands covered by Keck.}
\label{fig:Xray_spectra}
\end{figure*}

Another condition is that the prompt disk wind has to be transparent to X-rays on day 212. The wind has traveled out to a distance of $r_{\rm w}\approx 7\times 10^{16} {\rm cm}\ (\bar{v}_{\rm out}/0.12c)$. Assuming solar abundance, an order-of-magnitude estimate of the hydrogen column density is
\begin{eqnarray}
N_{\rm H}&\approx& \frac{X_{\rm H}M_{\rm w}/m_p}{4\pi r_{\rm w}^2} \nonumber \\
&\sim& 5\times 10^{21}\ {\rm cm^{-2}} \left(\frac{X_{\rm H}}{0.7}\right) \left(\frac{M_\mathrm{w}}{0.8M_\odot}\right)\left(\frac{r_\mathrm{w}}{10^{17}{\rm cm}}\right)^{-2}
\end{eqnarray}
where $X_{\rm H}$ is the hydrogen mass fraction and $M_{\rm w}$ is the mass of the wind.

We use the spectral synthesis code CLOUDY \citep{Ferland17} to simulate absorption by a wind of $M_\mathrm{w} =0.8M_\odot$, with uniform density out to $7\times 10^{16}$ cm. For the incident X-ray emission we adopt a total luminosity of $10^{41}\ {\rm erg\ s^{-1}}$. The spectrum is composed of two components with equal weight, a 0.1 keV blackbody and plateau at higher energies up to 10 keV, with power-law cutoffs at $0.1$ keV and $10$ keV. The shape in the range $0.1$--$10$ keV is similar to the observed spectrum of ULXs \citep{Kaaret17}. The result, shown in the left panel of Figure \ref{fig:Xray_spectra}, confirms that X-rays in the energy range of interest almost fully escape.

\subsection{Late Optical Emission by Reprocessing of X-rays}
\label{sec:late_optical}
As noted in Section \ref{sec:early_optical}, the late optical emission observed by Keck requires a different mechanism from the cooling emission of the disk wind. Here we consider the possibility of whether X-rays from the central BH invoked in the previous section can consistently explain the optical emission through reprocessing.

The X-rays from the center may partially irradiate the slow ejecta near the equatorial plane. This ejecta expand homologously with typical radius
$r_{\rm ej}= v_{\rm ej}t \sim 5\times 10^{14}\ {\rm cm} (v_{\rm ej}/10^3\ {\rm km\ s^{-1}})(t/60\ {\rm days})$. The hydrogen column density along the line of sight from the central BH to the ejecta is
\begin{eqnarray}
N_{\rm H}&\approx& \frac{X_{\rm H}M_{\rm ej}/m_p}{4\pi f_\Omega r_{\rm ej}^2} \nonumber \\
&\sim& 7\times 10^{23}\ {\rm cm^{-2}}f_\Omega^{-1} \left(\frac{X_{\rm H}}{0.7}\right)\left(\frac{M_{\rm ej}}{0.01M_\odot}\right)\left(\frac{r_{\rm ej}}{10^{15}\ {\rm cm}}\right)^{-2}
\end{eqnarray}
which is more than enough to fully absorb the soft X-rays. The X-rays are expected to thermalize in the ejecta and gets re-radiated in the optical, which may explain the late time optical emission observed on day $61$. 

In the case that the ejecta can efficiently thermalize the incident X-ray emission, the reprocessed emission can be approximated as blackbody emission with radius $r_{\rm ej}$ and luminosity $f_\Omega L_X$, where we adopt $L_X\approx 1.3\times 10^{41}\ {\rm erg\ s^{-1}}$ from Figure \ref{fig:early_optical}. We find that the luminosity integrated over the Keck bandwidth is consistent with the observed value $1.9\times 10^{40}\ {\rm erg\ s^{-1}}$ for $f_\Omega\approx 0.5$.

Using CLOUDY, we explore the lower limit on $M_{\rm ej}$ that reproduces the observed luminosity through reprocessing. The ejecta are assumed to be solar abundance and have a uniform density distribution, with an inner and outer edge of $10^{14}$ cm and $5\times 10^{14}$ cm respectively. Adopting the same incident spectrum as done for the disk wind case, we find that the observed luminosity can be reproduced for $M_{\rm ej}=0.01M_\odot$ if $f_\Omega\sim 1$, as shown in the right panel of Figure \ref{fig:Xray_spectra}. The temporal evolution for $t>50$ days is shown as a red dotted line in Figure \ref{fig:early_optical}. Overall, the required mass range for the slow ejecta of $M_{\rm ej}\gtrsim 0.01M_\odot$ is consistent with the theoretical prediction of mass ejection from BSGs.

If the slow ejecta can expand to the polar direction and increase its covering fraction to $f_{\rm ej}(>f_\Omega)$, one can additionally expect interaction of the slow ejecta and the outflow from the fallback disk still accreting at super-Eddington. The energy budget of this interaction is roughly
\begin{eqnarray}
&&\frac{1}{2}f_{\rm ej}\times({\rm outflow\ rate})\times \bar{v}_{\rm out}^2 \nonumber \\
&\sim &\frac{1}{2}f_{\rm ej}^2f_{\dot{M}}\dot{M}\bar{v}_{\rm out}^2 \nonumber \\
&\sim& 6\times 10^{41}\ {\rm erg\ s^{-1}}\left(\frac{f_{\rm ej}}{0.5}\right)^2\left(\frac{f_{\dot{M}}}{0.3}\right)\left(\frac{\bar{v}_{\rm out}}{0.12c}\right)^2\left(\frac{t}{61{\rm day}}\right)^{-5/3},
\end{eqnarray}
where we again assumed $\dot{M}\sim 10^{-8}M_\odot\ {\rm s^{-1}}(t/10^6\ {\rm sec})^{-5/3}$, and $f_{\dot{M}}$ is the fraction of the accreted mass that escapes as outflow. For a reprocessing efficiency similar to the above examples, this may be able to reproduce the optical emission. A similar dissipation/reprocessing mechanism was raised as one of the possibilities to explain the late-time optical emission of AT 2018cow \citep{Margutti19}. Hydrodynamical simulations of the ejecta and wind would be needed to clarify whether this mechanism is actually at work.

\section{Conclusion}
\label{sec:conclusion}
In this work we showed that the multi-wavelength observations of the recent transient AT 2018lqh reported by \cite{Ofek21} can be consistently explained by a BH with $\sim 30\,M_\odot$ forming from a rotating BSG star. Multi-dimensional radiation-hydrodynamical simulations would be an important future work to clarify the interplay of the ejecta and the disk and improve our simple modelling. Multi-wavelength follow-up, especially around tens of days that was unexplored for AT 2018lqh, is also key for validation of our model. 

Late X-ray observations can also be an important test. Assuming that the relation $\dot{M}\propto t^{-5/3}$ holds, we expect the accretion onto the BH to be super-Eddington for decades \citep{Tsuna20}. Thus continued observations in the X-ray may be a viable test that can support or falsify our interpretation of the late-time optical emission. Because of the presence of a radio point source 3" away from optical position, observations with resolution less than 3" is required to unambiguously identify the X-ray counterpart. We therefore suggest observations to be carried out by telescopes with high angular resolution, such as Chandra \citep{Weisskopf02}. However if the disk has already faded close to the Eddington luminosity, then X-rays of flux $F_X\approx 1.4\times 10^{-15}{\rm erg\ s^{-1}\ cm^{-2}} (L_X/10^{40}{\rm erg\ s^{-1}})$ may be marginal to detect even by Chandra.

If our interpretation is correct, the inferred event rate of AT 2018lqh ($1\%$ of the supernova rate) implies that transients with a BH and a disk are not so rare. This rate is between those of failed supernovae ($\sim 10\%$; \citealt{Adams17}) and gamma-ray bursts ($\sim 0.1\%$; e.g. \citealt{Butler10}). This is consistent with the naive expectation that failed supernovae come from stars with relatively slow rotation whereas gamma-ray bursts come from those with extreme rotation, and what we observed is the middle of these two. As disk wind generally shines on a timescale of days, we expect that optical/UV surveys with sub-day cadence, such as ZTF, ULTRASAT \citep{Sagiv14}, or Tomo-e Gozen \citep{Sako18} would observe more events like AT 2018lqh.

\begin{acknowledgements}
D.T. thanks Kohta Murase and the anonymous referee for helpful comments. D.T. is supported by the Advanced Leading Graduate Course for Photon Science (ALPS) at the University of Tokyo, and by the JSPS Overseas Challenge Program for Young Researchers. This work is also supported by JSPS KAKENHI Grant Numbers JP19J21578, JP20H05639, JP20K04010, JP20H01904, MEXT, Japan.
\end{acknowledgements}

\bibliography{references}
\bibliographystyle{aasjournal}

\end{document}